# Sterile neutrinos, dark matter and Standard Model extended by right-handed neutrinos

A.P. Serebrov[a,*], R.M. Samoilov[a], O.M. Zherebtsov[a] and N.S. Budanov[a]

NRC „Kurchatov Institute" – Petersburg Nuclear Physics Institute, Orlova roscha 1, Gatchina, Russia

e-mail: serebrov_ap@pnpi.nrcki.ru

**Abstract.** Joint analysis of the results of the Neutrino-4 experiment and the data of the GALLEX, SAGE and BEST experiments confirms the parameters of neutrino oscillations declared by the Neutrino-4 experiment ($\Delta m_{14}^2 = 7.3$ eV$^2$ and $\sin^2 2\theta_{14} \approx 0.36$) and increases the confidence level to $5.8\sigma$. Such a sterile neutrino thermalizes in cosmic plasma, contributes 5% to the energy density of the Universe, and can explain 15-20% of dark matter. It is discussed that the extension of the neutrino model by introducing two more heavy sterile neutrinos in accordance with the number of types of active neutrinos but with very small mixing angles to avoid thermalization will make it possible to explain the large-scale structure of the Universe and bring the contribution of sterile neutrinos to the dark matter of the Universe to the level of 27%. Sterile neutrinos are essentially right-handed neutrinos. The dynamic process of the dark matter generation, consisting of three right-handed neutrinos, is presented. Expansion of the Standard Model by introducing right-handed neutrinos seems possible. The scheme of masses and the scheme of mixing flavors are considered.

## 1. Introduction

There are quite a few indications of the possibility of the existence of a sterile neutrino. Anomalies were observed in several accelerator and reactor experiments: LSND at a confidence level of 3.8 σ [1], MiniBooNE 4.7 σ [2], reactor anomaly (RAA) 3σ [3,4], as well as in experiments with radioactive sources GALLEX/GNO, SAGE (gallium anomaly - GA 3.2σ) and BEST [5-7]. A detailed comparison of the results of the Neutrino-4 experiment [8] with the results of other experiments is presented in our paper [9]. Here we analyze the result of the Neutrino-4 experiment in connection with the possible role of sterile neutrinos in cosmology. In our previous work [10], the question of cosmological restrictions on sterile neutrinos was raised. In this work, we try to answer the previously posed questions. But we should start by presenting the result of the Neutrino-4 experiment and jointly analyzing the results of the Neutrino-4 experiment and the data from the GALLEX, SAGE, and BEST experiments. The result of the Neutrino-4 experiment is presented in Fig. 1.

A comparison of the results of the Neutrino-4 experiment and the results of the BEST experiment is presented in Fig. 2 on the left. A joint analysis was performed based on the data of the GALLEX, SAGE, and BEST experiments published in [7]. Obtained distribution $\Delta\chi^2(\Delta m_{14}^2, \sin^2 2\theta_{14})$ using this result together with the result of the Neutrino-4 experiment is shown in Fig.2 on the left. The parameter value at the best fit point is $\sin^2 2\theta_{14} = 0.38$, $\Delta m_{14}^2 = 7.3 \text{eV}^2$. The confidence level of the observation of oscillations obtained as a result of the joint analysis of the data was 5.8σ [9].

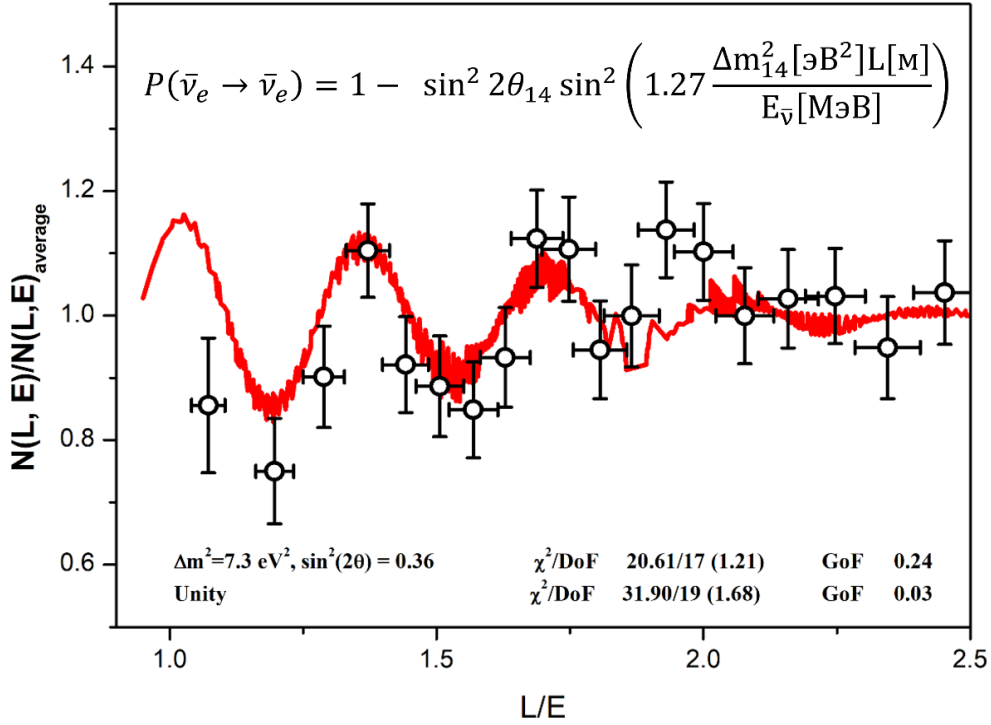

Fig.1. Oscillation dependence. Oscillation parameters obtained from data analysis are $m_{14}^2 = \left(7.3 \pm 0.13_{st} + 1.16_{sys}\right)$ eV$^2$ $\sin^2 2\theta_{14} = 0.36 \pm 0.12_{stat}$ (2.9$\sigma$)

However, there are several experiments (STEREO [11], PROSPECT [12], DANSS [13], NEOS+RENO [14], RAA [4]) aimed at searching for sterile neutrinos, in which the effect of oscillations was not detected. Therefore, it is necessary to understand the reasons for this situation (Fig. 2 on the right). In principle, one must ask which results should be preferred: the results of direct experiments or the results obtained during the calculation of complex processes? It should be noted that the Reactor Antineutrino Anomaly (RAA) is based on a rather complex method of absolute measurements, but the Neutrino-4 experiment uses the method of relative measurements and does not require accurate knowledge of the spectrum of reactor antineutrinos. The BEST experiment uses the well-known spectrum of monochromatic neutrinos and is also more reliable. It can be said that the discrepancy between the Neutrino-4 result, and the RAA is a discrepancy between direct relative measurements and absolute measurements that depend on complex reactor calculations.

The Neutrino-4 experiment is aimed at directly measuring the oscillation parameters $\sin^2 2\theta_{14}$ and $\Delta m_{14}^2$. The experiment used the method of relative measurements and the method of coherent summation of measurement results for different neutrino energies to avoid the effect of averaging oscillations over the spectrum. As a result of the experiment, it was found that the oscillation period for an average energy value of 4 MeV is 1.5 m. The biological protection of research reactors is 5-6 m, and at nuclear plants - 10 m. Therefore, outside the biological protection for research reactors, the oscillation effect is averaged over the spectrum, however, there is a deficit in the total neutrino flux, which is equal to $0.5 \sin^2 2\theta_{14}$. In experiments at nuclear power plants, the effect of oscillations is averaged even in the reactor core, because the size of the active zone is approximately 4 m. Therefore, for the experiments DANSS, NEOS+RENO, the sensitivity is not enough for measurements in the region $\sin^2 2\theta_{14} = 0.38$, $\Delta m_{14}^2 = 7.3$ eV$^2$. The only option left is to measure the neutrino flux deficit by using absolute measurements and precise calculations of the complex neutrino spectrum from approximately 700 isotopes. The discrepancy between $\sin^2 2\theta_{14}$ from the RAA and the estimate $\sin^2 2\theta_{14} \approx 0.35^{+0.09}_{-0.07}$ from the Neutrino-4 experiment and GA is 2.4σ. Although this discrepancy has not yet gone beyond 3σ, a possible interpretation of this situation is required.

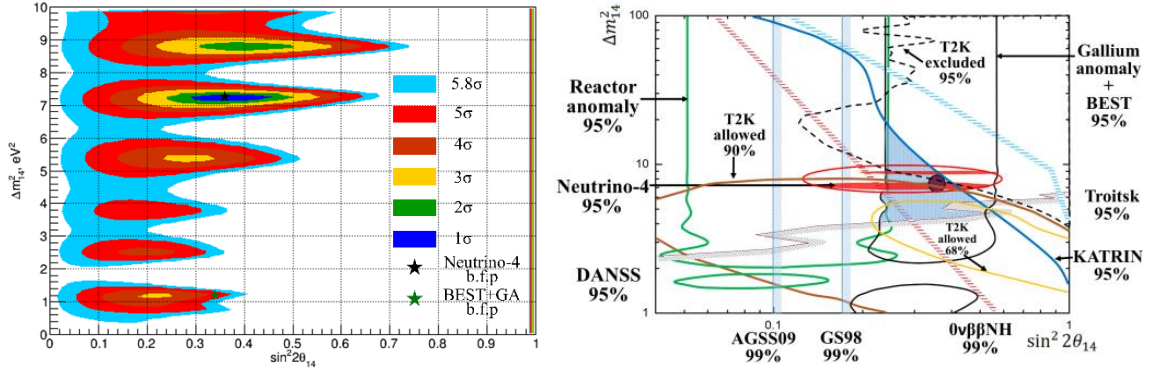

Fig. 2. On the left – The result of a joint analysis of GA, BEST and Neutrino-4, where blue indicates the area with a confidence of 1σ, green - 2σ, yellow - 3σ, dark red - 4σ, red - 5σ and light blue - 5.8σ. On the right – the comparison of the results of the Neutrino-4 experiment with the results of the KATRIN and GERDA experiments. Exclusion contours are from [15] and [16].

Calculating the absolute power of a reactor with high accuracy is a big challenge. For example, the energy carried away by antineutrinos is 5%, but there is no indication in publications that this effect was considered. In addition, the residual power of the reactor immediately after the reactor is turned off is 5%, but this issue is also not discussed. Finally, there appear to be unaccounted beta decays with a short lifetime and high decay energy because there is a distortion of the spectrum in relation to the calculated spectrum. It gives a deficit of about 5%. This is the so-called "bump" in the energy region of 5 MeV, which arises because the ratio of the experimental spectrum to the calculated spectrum is used, and the ratio of the calculated spectrum to the experimental spectrum should be analyzed. Then, this will be a hole, which indicates that short-lived isotopes with high decay energy are not taken into account.

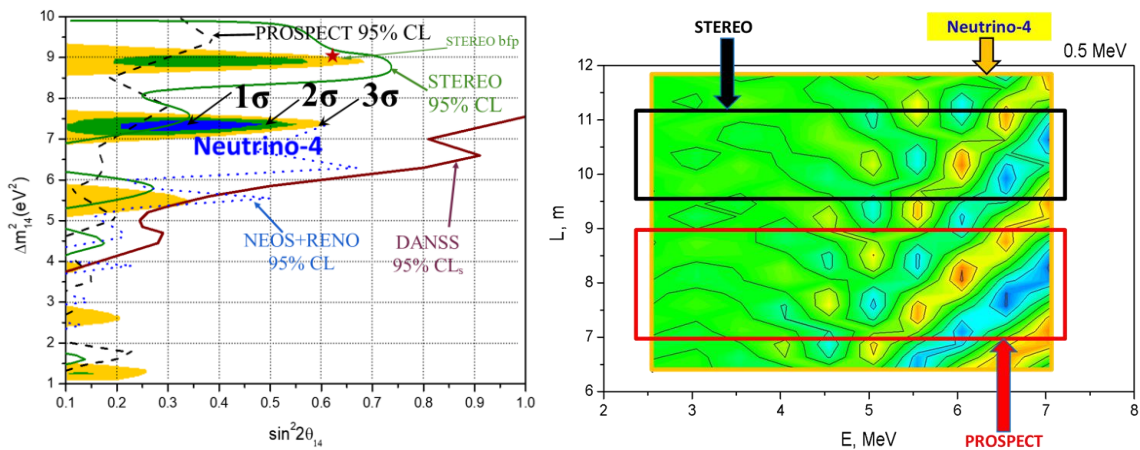

Fig. 3. On the left – comparison of the sensitivity of the experiments Neutrino-4, STEREO, PROSPECT, DANSS and NEOS. On the right – comparison of regions (L, E) in measurements of the experiments: Neutrino-4, STEREO and PROSPECT.

A comparison of the results of the Neutrino-4 experiment with the results of the PROSPECT, STEREO experiments shows that without the method of coherent summation of measurement results at different neutrino energies, it is impossible to observe oscillations with a period of 1.5 m. The fact is that for the method of coherent summation of measurement results at different neutrino energies it is required to make measurements over a wide range of distances, as was done in the Neutrino-4 experiment, and to sum the measurement results with the same L/E ratio, i.e. with the same oscillation phase.

As can be seen from Fig. 3, the range of measurements at different distances in the PROSPECT and STEREO experiments is significantly smaller than in the Neutrino-4 experiment; therefore, it is not possible to distinguish the effect of oscillations with a period of 1.5 m. We suppose that the STEREO and PROSPECT experiments, in order to correctly compare their own results with the results of the Neutrino-4 experiment, should present the data in the form of the L/E dependence.

It is of interest to compare the results of the KATRIN and GERDA experiments with the results of the Neutrino-4 experiment (Fig. 2 on the right). The result of the KATRIN experiment does not exclude the Neutrino-4 region for $\sin^2 2\theta_{14} \leq 0.4$ [15]. Data on restrictions on oscillation parameters related to consequences from the results of experiments on double neutrinoless beta decay are taken from [15]. The GERDA experiment [17] requires special attention since it is aimed at searching for the mass of Majorana-type neutrinos. Currently, the strongest limit on Majorana mass obtained from the GERDA experiment is 3 times less than the Majorana mass prediction obtained from the Neutrino-4 experiment. If in the future the Majorana mass limit of the double beta decay experiment is lowered and the result of the Neutrino-4 experiment is confirmed, this will close the light Majorana neutrino hypothesis.

In general, from the presented analysis we can conclude that a sterile neutrino with a mass of 2.7 eV is, at a minimum, not excluded by other experiments and is confirmed by the BEST experiment. Now, let's move on to the question of the role of sterile neutrinos in cosmology.

## 2. The role of sterile neutrinos in cosmology, Standard model expanding by introducing right-handed neutrinos

As shown in [10] sterile neutrinos with $\sin^2 2\theta_{14} = 0.38, \Delta m_{14}^2 = 7.3 \text{eV}^2$ thermalized in primordial plasma and their density is the same as the density of active neutrinos. In this work, we took a closer look at previous estimates of the frequency of neutrino collisions with each other and found that the effect of inelastic scattering was not taken into account.

The fact is that any neutrino-neutrino scattering leads to a change in the neutrino energy and the process of coherent propagation of neutrinos in the medium is interrupted. The imaginary part of the scattering amplitude includes absorption and scattering with energy changes. It was the scattering with changes in energy that was not considered. Taking this effect into account increases the collision frequency by approximately 120 times. An increase in the collision frequency leads to an increase in the rate of thermalization of sterile neutrinos. Here are the results of calculations taking this increasing into account.

Based on the sterile neutrino mass $m_{\nu_4} = 2.7\text{eV}$ the contribution of such sterile neutrinos to the energy density of the Universe was estimated. It amounted to 5% of the total density or 15-20% of the dark matter density. To explain the entire density of dark matter, two more heavy sterile neutrinos can be introduced into consideration in accordance with the number of types of active neutrinos, but with very small mixing angles to avoid thermalization. This makes it possible to bring the contribution of sterile neutrinos to the dark matter of the Universe to the level of 27% and explain the large-scale structure of the Universe. Considering the effect of increasing the collision frequency led to an even greater decrease in the mixing angle of heavy sterile neutrinos. Taking this circumstance into account fig. 4 (left) shows the possible values $\Delta m^2, \sin^2 2\theta$, which can contribute to the density of the Universe 25-20% (red area), 20-15% (yellow area), 15-10% (green area) and 10-5% (blue area). The same fig. 4 (left) shows the regions of restrictions on sterile neutrinos from astrophysical and laboratory experiments.

Now it should be clarified that sterile neutrinos are essentially right-handed neutrinos. Therefore, the introduction of sterile neutrinos into consideration means only minimal extension of the Standard Model by right-handed neutrinos. Figure 4 (right) illustrates the scheme for extending the Standard Model by introducing additional elementary particles. This approach to the problem of dark matter means that dark matter can be explained in terms of an Extended Standard Model with right-handed neutrinos. It should be noted that there is the so-called Neutrino Minimal Standard Model $\nu$MSM [18], in which the existence of keV-scale Majorana neutrinos is assumed. However, we are discussing an extension of the model by right-handed Dirac neutrinos. Figures 5 and 6 (left) illustrate the schemes if mixing of the left- and right-handed neutrinos and antineutrinos and fig. 6 (right) shows matrices of this mixing.

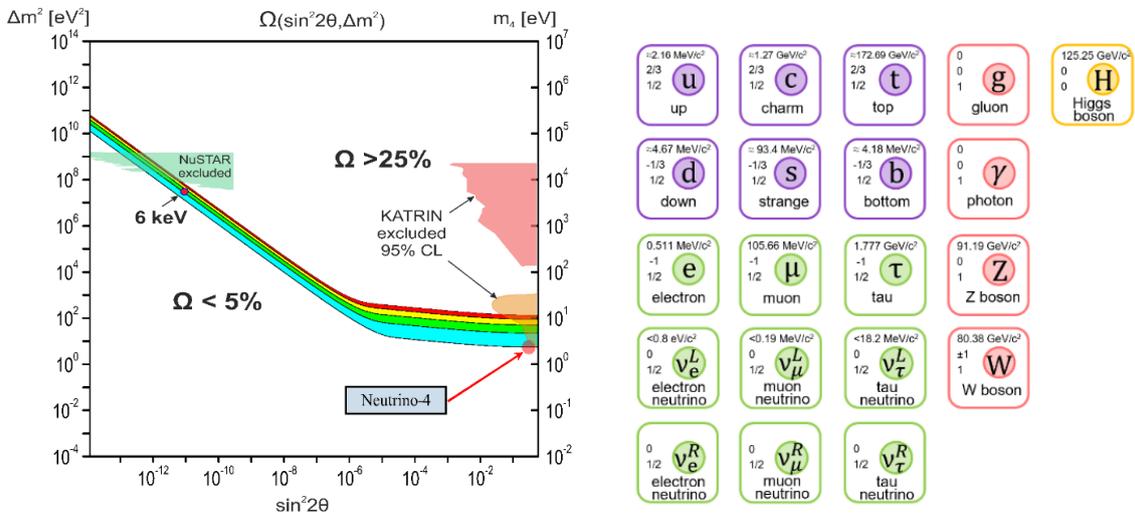

Fig. 4. On the left – Laboratory and astrophysical constraints on the parameters of sterile neutrinos. Red spots – result of the Neutrino-4 experiment and possible mass of the heavy right-handed neutrinos; green area – restrictions from the NuSTAR experiment [20]; orange area – KATRIN excluded 95% CL – restrictions from the KATRIN experiment for sterile neutrinos with a mass of ~ 1 eV [15]; red area – 95% CL limitations from experiments on measuring the mass of electron neutrinos from [21]; On the right – Scheme of Extended Standard Model by introducing additional elementary particles - right-handed Dirac neutrinos.

If we assume that the mass of the light right-handed neutrino is determined, then the masses of heavy right-handed neutrinos are unknown. It is interesting to understand in what range these masses are and whether they are already excluded by experimental astrophysical or laboratory data? Figure 4 shows the possible mass of heavy right-handed neutrinos. It can be seen the range is practically unlimited up to 6 keV. Figure 7 presents astrophysical restrictions on the region of masses and mixing angles. The red ellipse covers part of the region of possible masses and mixing angles of right-handed neutrinos, which are not yet excluded by direct astrophysical observations.

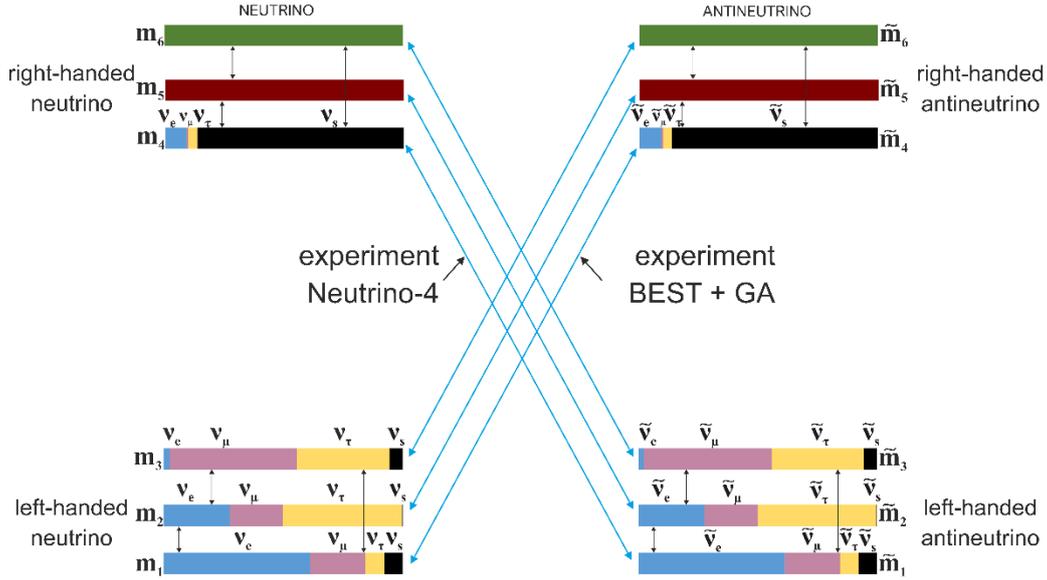

Fig. 5. Scheme of masses in the left-right neutrino model. Arrows indicate possible mixing between mass states. It is important to note that mixing occurs between right-handed antineutrinos and left-handed neutrinos, right-handed neutrinos and left-handed antineutrinos, respectively. Mixings related to the Neutrino-4 experiment and the BEST and GA experiment are shown. For all cases, a direct mass hierarchy is assumed, although this does not exclude an inverse mass hierarchy, both for left-handed neutrinos and right-handed neutrinos.

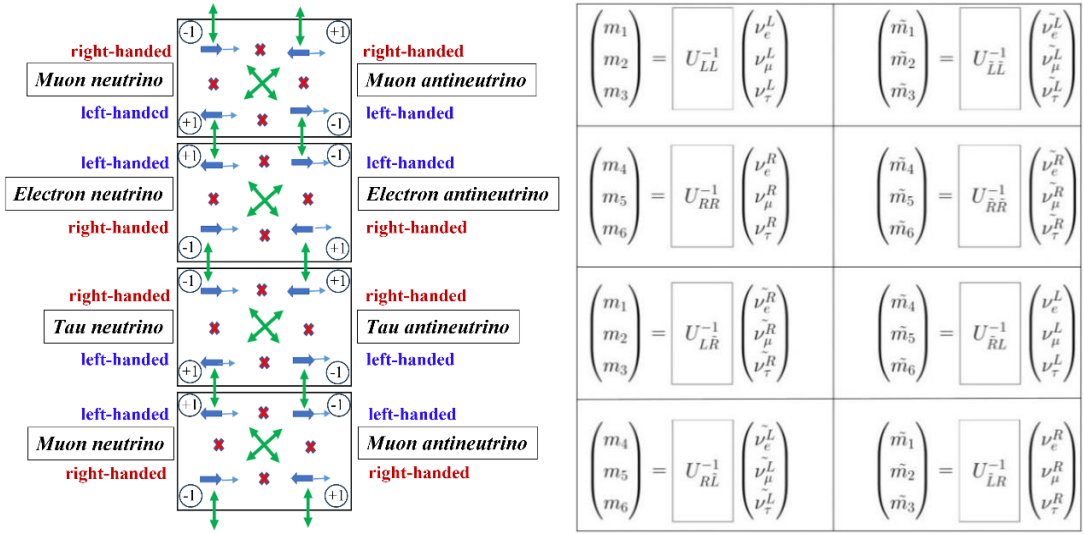

Fig. 6. On the left – scheme of flavor mixing in the left-right neutrino model. Red crosses indicate the absence of flavor mixing. Green arrows indicate possible flavor combinations. Allowed mixing is determined by the same chirality of states. Forbidden mixings are determined by the opposite chirality of states. Thus, in this model the neutrino lepton number is conserved. The lepton number is positive for left-handed neutrinos and negative for left-handed antineutrinos, while the lepton number is negative for right-handed neutrinos and positive for right-handed antineutrinos as indicated in the corners of the rectangles. On the right – matrices that define the unitary transformation of flavor states to mass states. The inverse transformation from mass states to flavor states is carried out through a transposed and complex conjugate matrix, what is usually called the PMNS matrix. Mass states have a certain mass, the same for neutrinos and antineutrinos. The complete 12x12 mixing matrix, consisting of the above matrices, is indicated in the appendix.

It should be noted that the Lyman-α constraints, indicated by the light brown dashed line in the figure, are essentially model-dependent. To obtain these constraints, a comparison of direct astrophysical observations with the results of simulations is used. Simulations are made under the assumption of one or another cosmological model of dark matter. In addition, the observed shape of the flux power spectrum can be affected by factors which contribution has not yet been correctly assessed, for example, Doppler temperature broadening [22]. Moreover, in some considerations, the observed spectrum is consistent with both model of cold dark matter and of warm dark matter, which can be formed by a sterile neutrino with a mass of ~ 1 keV [23].

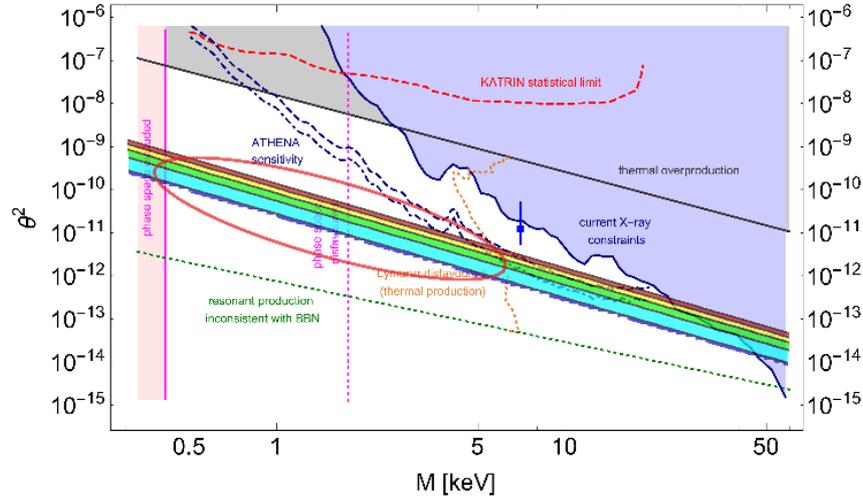

Fig. 7. Comparison of allowable values of mixing parameters obtained as a result of calculations with existing constraints from astrophysical observations. On the left is data on astrophysical boundaries in 2018 from [22].

### 3. Dynamics of dark matter generation and stability of dark matter consisting of right-handed neutrinos

The next important problem to be discussed is the lifetime of right-handed neutrinos. It was shown in [24–26] that the decay of right-handed neutrinos is possible through the channels of two-particle and three-particle decay. Fig. 8 shows the decay time depending on the neutrino mass. The vertical axis shows the ratio of the lifetime to the age of the Universe - 13.8 billion years. Decay time of the 1 MeV sterile neutrinos is equal to the age of the Universe. Thus, right-handed neutrinos with a mass of 1 MeV or more are no longer suitable for dark matter.

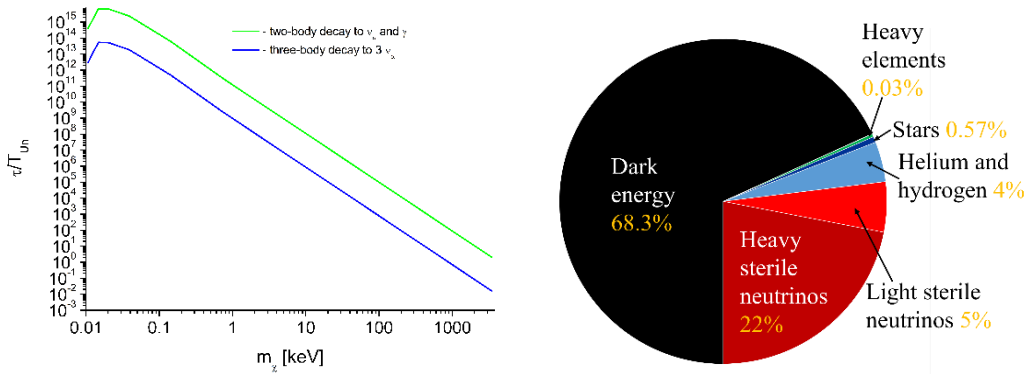

Fig. 8. On the left – Decay time of right-handed neutrinos in the channel of two-body and three-body decay. The lifetimes are reduced to the age of the Universe. The mass and mixing angle of a sterile and electron neutrino and correspond to the condition that the fraction of dark matter from such neutrinos is 17%. On the right – General picture of the composition of the energy density and mass of the Universe.

Three-body decay is main decay mode, and two-body decay has a probability two orders of magnitude smaller. However, two-body decay is of great importance because monochromatic gamma quanta appear with an energy equal to half the mass of a heavy neutrino. Since the mass of an active neutrino is extremely small, the energies of a gamma quantum and an active neutrino are practically equal. The presence of such monochromatic radiation is a verification for the existence of heavy neutrinos with the corresponding mass. The difficulty is in the fact that less than 10 keV neutrinos has the lifetime that exceeds the age of the Universe by 8 orders, and 10 eV neutrinos has the lifetime 12 orders longer than the age of the Universe. That is why the region of experimental limitations ends at a mass of 6 keV. It is interesting to note that the observation of monochromatic gamma rays with an energy of 3.5 keV corresponding to a heavy neutrino mass of 7 keV was reported in [27]. However, this observation was subsequently disfavored. Thus, the masses of heavy right-handed neutrinos less than 6 keV have not yet been disfavored, therefore, the mass of the heaviest right neutrino is $m_{\nu_\tau^R} < 6$ keV. However, a heavy right-handed neutrino with a mass less than 6 keV is still of interest as a particle of warm dark matter. Thus, dark matter with right-handed neutrino masses $m_{\nu_\tau^R} < 6$ keV is quite stable because the decay time is 10-11 orders of magnitude longer than the existence of the Universe.

The dynamics of the generation of dark matter, consisting of three right-handed neutrinos, is shown in Fig. 9. New calculations taking into account the effect of inelastic scattering are presented.

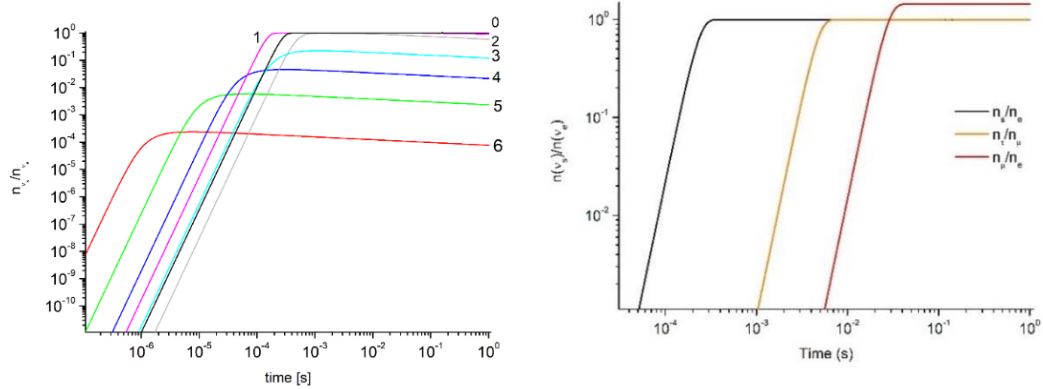

Fig. 9. On the left – Dynamics of the generation of the dark matter consisted of the right-handed neutrinos. Line 0 is for the oscillation parameters values $\sin^2 2\theta = 0.36, m = 2.7$ eV, line 1 - $\sin^2 2\theta = 7 \cdot 10^{-3}, m = 14.1$ eV, line 2 - $\sin^2 2\theta = 1 \cdot 10^{-5}, m = 20$ eV, line 3 - $\sin^2 2\theta = 3.7 \cdot 10^{-8}, m = 0.17$ keV, line 4 - $\sin^2 2\theta = 8 \cdot 10^{-10}, m = 1.4$ keV, line 5 - $\sin^2 2\theta = 1 \cdot 10^{-11}, m = 14$ keV, line 6 - $\sin^2 2\theta = 1.6 \cdot 10^{-14}, m = 360$ keV.
On the right – Process of neutrinos decoupling.

The heaviest right-handed neutrino decouples from the plasma first and with the smallest mixing angle with the left-handed neutrinos. The less heavy right-handed neutrino decouples from the plasma later at a larger mixing angle, which is also small so that there is no thermalization. Previously decoupled heavy right-handed neutrinos are already beginning to form gravitational regions, since at times $10^{-5}$ -$10^{-4}$ s, the density of dark matter particles is quite high and the law of gravitational attraction $1/r^2$ works quite effectively. Heavy sterile (right-handed) neutrinos with a mass of several keV can form structures not only due to gravitational forces, but also due to the forces of attraction between right-handed neutrinos and right-handed antineutrinos.

Light right-handed neutrinos with mass $m_{\nu_e^R} = 2.7$ eV become non-relativistic upon transition to the matter epoch (80 thousand years), because by this time, the plasma temperature is 0.26 eV. At this stage, light right-handed neutrinos combine with dark matter from heavy right-handed neutrinos.

**4. Constraints on the sum of masses of relic neutrinos**

There are boundaries on the sum of the masses of relic neutrinos ~0.12 eV. However, right-handed light neutrinos are non-relativistic at the matter epoch of the expansion of the Universe (80 thousand years) and they are captured by dark matter formed by heavy right-handed neutrinos. They are captured by dark matter from right-handed heavy neutrinos just like other non-relativistic particles. Therefore, there are no relic neutrinos with a mass of 2.7 eV. All left-handed neutrinos are relic cosmic background radiation, and all right-handed neutrinos are part of dark matter. Such a scheme can be logically considered and requires specific calculations.

**5. Experimental constraints on light sterile neutrino from astrophysical data on measuring the mass content of $^4$He.**

Light sterile neutrino, unlike heavy neutrinos, it is thermalized in cosmic plasma and affects the rate of expansion of the plasma. This issue requires special consideration. Light right-handed neutrinos affect the thermodynamic process, since are still relativistic and give an additional degree of freedom, which affects the rate of expansion and, accordingly, the rate of plasma cooling. This leads to the production of more $^4$He than in the case of only three left-handed neutrinos.

There is a constraint on the number of neutrino types from astrophysical measurements of the $^4$He abundance. The calculation results are illustrated in Fig. 10 on the left, based on the data from [28,29]. For further analysis, we consider the results of four measurements of the mass fraction of $^4$He: Izotov 2014 ($Y_P = 0.2551 \pm 0.0022$) [30], Aver 2015 ($Y_P = 0.2449 \pm 0.0040$) [31], EMPRESS 2022 ($Y_P = 0.2370^{+0.0034}_{-0.0033}$ [32]) and Tsivilev 2023 ($Y_P \geq 0.2493 - 0.2940$) [33], which characterize the spread of measurement results. In addition, [34] gives a detailed analysis of the results of all $Y_P$ measurements (except for the last result of EMPRESS 2022), as well as the result of the analysis of astrophysical observations, which presents the value $Y_P = 0.2462 \pm 0.0022$. Kuricin 2022 [35] presents one of the most accurate values of $Y_P = 0.2470 \pm 0.0020$ at the moment.

These results are demonstrated in Fig. 10 on the right together with the calculated predictions of the $^4$He abundance from the baryon asymmetry of the Universe and the neutron lifetime to compare the calculated predictions of the $^4$He abundance in the $N_\nu = 3$ and $N_\nu = 4$ models with the results of experimental measurements. The measurement results of Aver 2015 and Kurichin 2022 are in good agreement with the $N_\nu = 3$ model, and the results of Izotov 2014 are closer to the prediction of the $N_\nu = 4$ model. Finally, the latest EMPRESS 2022 as well as Tsivilev 2023 results do not fit into any model.

Thus, based on the presented astrophysical data, it is impossible to draw a definite conclusion in favor of the model of three or four neutrinos. In any case, one cannot conclude that cosmology, based on astrophysical data, forbids sterile neutrinos with parameters $\Delta m_{14}^2 = 7.3$ eV$^2$ and $\sin^2 2\theta_{14} = 0.36$.

In addition, there is the issue of lepton asymmetry, which can affect the ratio of neutrons to protons. At the beginning of BBN, neutrons and protons are in equilibrium until the equilibrium is disturbed by a weak interaction. If the $p + \bar{v}_e \to n + e^+$ process is suppressed with respect to the $n + v_e \to p + e^-$ process due to a smaller number of electron antineutrinos, then this suppresses the neutron-proton ratio and, as a result, $Y_P$ decreases. This decrease in $Y_P$ can be compensated by increasing the number of degrees of freedom $N_v^{\text{eff}}$ to keep the same value of $Y_P$. Thus, the presence of lepton asymmetry disguises the presence of the fourth neutrino. Therefore, the question of the role of neutrino-antineutrino asymmetry requires special consideration.

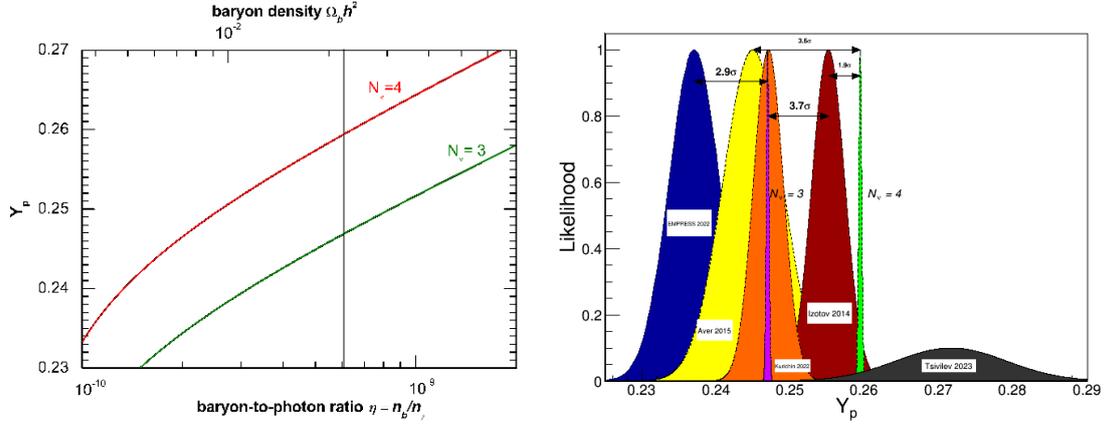

Fig. 10. On the left - $Y_p$ abundances as a function of baryon asymmetry at $N_v$ = 3 and 4 respectively. The line thickness is determined by the experimental accuracy of measuring the neutron lifetime ($\tau_n = 879.4 \pm 0.6$ s). The vertical line corresponds to the value of the baryon asymmetry (6.090 ± 0.060)·10$^{-1}$, and its thickness corresponds to one standard deviation. Data taken from [27] On the right - Comparison of the calculated predictions of the abundance of 4He with the known neutron lifetime and the value of the baryon asymmetry in the model $Nv$ = 3 and $Nv$ = 4 (purple and green peaks, respectively) with the results of astrophysical observations: Izotov 2014, Aver 2015, Kurichin 2022, EMPRESS 2022 and Tsivilev 2023 (red, yellow, orange, blue and gray distribution respectively).

**Conclusion.**

1. A brief analysis of the result of the Neutrino-4 experiment and the results of other experiments in the search for sterile neutrinos is presented. A joint analysis of the results of the Neutrino-4 experiment and data from the GALLEX, SAGE and BEST experiments confirms the parameters of neutrino oscillations declared by the Neutrino-4 experiment ($\Delta m_{14}^2 = 7.3\text{eV}^2$ and $\sin^2 2\theta_{14} \approx 0.36$) and increases the confidence level to 5.8σ. The estimated contribution to the energy density of the Universe from sterile neutrinos with these parameters is 5%.

2. Expanding the neutrino model by introducing two more heavy sterile neutrinos in accordance with the number of types of active neutrinos will make it possible to explain the structure of the Universe and bring the contribution of sterile neutrinos to the dark matter of the Universe to the level of 27%.

3. The dynamic process of the generation of dark matter, consisting of three right-handed neutrinos, is presented.

4. It is shown that, based on modern astrophysical data, it is impossible to make a definite conclusion in favor of the model of three or four neutrinos.

This work was supported by the Russian Science Foundation (Project No. 24-12-00091).

# Appendix

Full 12x12 matrix of left-right neutrino model. It converts 12 mass states into 12 flavor states.

$$\nu_L = \begin{pmatrix} \nu_e^L \\ \nu_\mu^L \\ \nu_\tau^L \end{pmatrix} \qquad \tilde{\nu}_L = \begin{pmatrix} \tilde{\nu}_e^L \\ \tilde{\nu}_\mu^L \\ \tilde{\nu}_\tau^L \end{pmatrix} \qquad \nu_R = \begin{pmatrix} \nu_e^R \\ \nu_\mu^R \\ \nu_\tau^R \end{pmatrix} \qquad \tilde{\nu}_R = \begin{pmatrix} \tilde{\nu}_e^R \\ \tilde{\nu}_\mu^R \\ \tilde{\nu}_\tau^R \end{pmatrix}$$

$$m_L = \begin{pmatrix} m_1 \\ m_2 \\ m_3 \end{pmatrix} \qquad \tilde{m}_L = \begin{pmatrix} \tilde{m}_1 \\ \tilde{m}_2 \\ \tilde{m}_3 \end{pmatrix} \qquad m_R = \begin{pmatrix} m_4 \\ m_5 \\ m_6 \end{pmatrix} \qquad \tilde{m}_R = \begin{pmatrix} \tilde{m}_4 \\ \tilde{m}_5 \\ \tilde{m}_6 \end{pmatrix}$$

$$U = \begin{pmatrix} U_{LL} & 0 & 0 & U_{L\tilde{R}} \\ 0 & U_{\tilde{L}\tilde{L}} & U_{\tilde{L}R} & 0 \\ 0 & U_{R\tilde{L}} & U_{RR} & 0 \\ U_{\tilde{R}L} & 0 & 0 & U_{\tilde{R}\tilde{R}} \end{pmatrix} \qquad \begin{pmatrix} \nu_L \\ \tilde{\nu}_L \\ \nu_R \\ \tilde{\nu}_R \end{pmatrix} = \begin{pmatrix} U_{LL} & 0 & 0 & U_{L\tilde{R}} \\ 0 & U_{\tilde{L}\tilde{L}} & U_{\tilde{L}R} & 0 \\ 0 & U_{R\tilde{L}} & U_{RR} & 0 \\ U_{\tilde{R}L} & 0 & 0 & U_{\tilde{R}\tilde{R}} \end{pmatrix} \begin{pmatrix} m_L \\ \tilde{m}_L \\ m_R \\ \tilde{m}_R \end{pmatrix}$$

$$\begin{pmatrix} \nu_e^L \\ \nu_\mu^L \\ \nu_\tau^L \\ \tilde{\nu}_e^L \\ \tilde{\nu}_\mu^L \\ \tilde{\nu}_\tau^L \\ \nu_e^R \\ \nu_\mu^R \\ \nu_\tau^R \\ \tilde{\nu}_e^R \\ \tilde{\nu}_\mu^R \\ \tilde{\nu}_\tau^R \end{pmatrix} = \begin{pmatrix} U_{e1}^{LL} & U_{e2}^{LL} & U_{e3}^{LL} & 0 & 0 & 0 & 0 & 0 & 0 & U_{e\tilde{1}}^{L\tilde{R}} & U_{e\tilde{2}}^{L\tilde{R}} & U_{e\tilde{3}}^{L\tilde{R}} \\ U_{\mu 1}^{LL} & U_{\mu 2}^{LL} & U_{\mu 3}^{LL} & 0 & 0 & 0 & 0 & 0 & 0 & U_{\mu\tilde{1}}^{L\tilde{R}} & U_{\mu\tilde{2}}^{L\tilde{R}} & U_{\mu\tilde{3}}^{L\tilde{R}} \\ U_{\tau 1}^{LL} & U_{\tau 2}^{LL} & U_{\tau 3}^{LL} & 0 & 0 & 0 & 0 & 0 & 0 & U_{\tau\tilde{1}}^{L\tilde{R}} & U_{\tau\tilde{2}}^{L\tilde{R}} & U_{\tau\tilde{3}}^{L\tilde{R}} \\ 0 & 0 & 0 & U_{\tilde{e}\tilde{1}}^{\tilde{L}\tilde{L}} & U_{\tilde{e}\tilde{2}}^{\tilde{L}\tilde{L}} & U_{\tilde{e}\tilde{3}}^{\tilde{L}\tilde{L}} & U_{\tilde{e}1}^{\tilde{L}R} & U_{\tilde{e}2}^{\tilde{L}R} & U_{\tilde{e}3}^{\tilde{L}R} & 0 & 0 & 0 \\ 0 & 0 & 0 & U_{\tilde{\mu}\tilde{1}}^{\tilde{L}\tilde{L}} & U_{\tilde{\mu}\tilde{2}}^{\tilde{L}\tilde{L}} & U_{\tilde{\mu}\tilde{3}}^{\tilde{L}\tilde{L}} & U_{\tilde{\mu}1}^{\tilde{L}R} & U_{\tilde{\mu}2}^{\tilde{L}R} & U_{\tilde{\mu}3}^{\tilde{L}R} & 0 & 0 & 0 \\ 0 & 0 & 0 & U_{\tilde{\tau}\tilde{1}}^{\tilde{L}\tilde{L}} & U_{\tilde{\tau}\tilde{2}}^{\tilde{L}\tilde{L}} & U_{\tilde{\tau}\tilde{3}}^{\tilde{L}\tilde{L}} & U_{\tilde{\tau}1}^{\tilde{L}R} & U_{\tilde{\tau}2}^{\tilde{L}R} & U_{\tilde{\tau}3}^{\tilde{L}R} & 0 & 0 & 0 \\ 0 & 0 & 0 & U_{e\tilde{1}}^{R\tilde{L}} & U_{e\tilde{2}}^{R\tilde{L}} & U_{e\tilde{3}}^{R\tilde{L}} & U_{e1}^{RR} & U_{e2}^{RR} & U_{e3}^{RR} & 0 & 0 & 0 \\ 0 & 0 & 0 & U_{\mu\tilde{1}}^{R\tilde{L}} & U_{\mu\tilde{2}}^{R\tilde{L}} & U_{\mu\tilde{3}}^{R\tilde{L}} & U_{\mu 1}^{RR} & U_{\mu 2}^{RR} & U_{\mu 3}^{RR} & 0 & 0 & 0 \\ 0 & 0 & 0 & U_{\tau\tilde{1}}^{R\tilde{L}} & U_{\tau\tilde{2}}^{R\tilde{L}} & U_{\tau\tilde{3}}^{R\tilde{L}} & U_{\tau 1}^{RR} & U_{\tau 2}^{RR} & U_{\tau 3}^{RR} & 0 & 0 & 0 \\ U_{\tilde{e}1}^{\tilde{R}L} & U_{\tilde{e}2}^{\tilde{R}L} & U_{\tilde{e}3}^{\tilde{R}L} & 0 & 0 & 0 & 0 & 0 & 0 & U_{\tilde{e}\tilde{1}}^{\tilde{R}\tilde{R}} & U_{\tilde{e}\tilde{2}}^{\tilde{R}\tilde{R}} & U_{\tilde{e}\tilde{3}}^{\tilde{R}\tilde{R}} \\ U_{\tilde{\mu}1}^{\tilde{R}L} & U_{\tilde{\mu}2}^{\tilde{R}L} & U_{\tilde{\mu}3}^{\tilde{R}L} & 0 & 0 & 0 & 0 & 0 & 0 & U_{\tilde{\mu}\tilde{1}}^{\tilde{R}\tilde{R}} & U_{\tilde{\mu}\tilde{2}}^{\tilde{R}\tilde{R}} & U_{\tilde{\mu}\tilde{3}}^{\tilde{R}\tilde{R}} \\ U_{\tilde{\tau}1}^{\tilde{R}L} & U_{\tilde{\tau}2}^{\tilde{R}L} & U_{\tilde{\tau}3}^{\tilde{R}L} & 0 & 0 & 0 & 0 & 0 & 0 & U_{\tilde{\tau}\tilde{1}}^{\tilde{R}\tilde{R}} & U_{\tilde{\tau}\tilde{2}}^{\tilde{R}\tilde{R}} & U_{\tilde{\tau}\tilde{3}}^{\tilde{R}\tilde{R}} \end{pmatrix} \begin{pmatrix} m_1 \\ m_2 \\ m_3 \\ \tilde{m}_1 \\ \tilde{m}_2 \\ \tilde{m}_3 \\ m_4 \\ m_5 \\ m_6 \\ \tilde{m}_4 \\ \tilde{m}_5 \\ \tilde{m}_6 \end{pmatrix}$$